\documentclass[12pt]{article}

\usepackage{mathptmx}
\usepackage{cite} % collapse multiple cites, eg: 1,2,3,4,5 -> 1-5
\usepackage[scaled=0.9]{helvet}
\usepackage{graphicx}
\usepackage{amsmath}

\usepackage{url}  % typeset url's
\usepackage{array}
\def\tstrut{\vrule height3.5ex depth0pt width0pt} % used in tables
\def\fom#1{F^{\mbox{\scriptsize Omn\`es}}_#1(q^2_#1,F_0^#1,F_1^#1,F_2^#1)}

\evensidemargin\oddsidemargin
\parskip=\medskipamount

\makeatletter
\long\def\@makecaption#1#2{%
  \vskip\abovecaptionskip
  \sbox\@tempboxa{\small{\bfseries #1} \  #2}%
  \ifdim \wd\@tempboxa >\hsize
    \small{\bfseries #1} \  #2\par
  \else
    \global \@minipagefalse
    \hb@xt@\hsize{\hfil\box\@tempboxa\hfil}%
  \fi
  \vskip\belowcaptionskip}
\renewcommand\section{\@startsection {section}{1}{\z@}%
      {-3.25ex\@plus -1ex \@minus -.2ex}%
      {1ex \@plus .2ex}%
      {\normalfont\large\sffamily\bfseries}}
\renewcommand\subsection{\@startsection{subsection}{2}{\z@}%
      {-3ex\@plus -1ex \@minus -.2ex}%
      {0.5ex \@plus .2ex}%
      {\normalfont\normalsize\sffamily\bfseries}}
\renewcommand\subsubsection{\@startsection{subsubsection}{3}{\z@}%
      {-3ex\@plus -1ex \@minus -.2ex}%
      {0.25ex \@plus .2ex}%
      {\normalfont\normalsize\sffamily\bfseries}}
\renewcommand\paragraph{\@startsection{paragraph}{4}{\z@}%
      {3ex \@plus1ex \@minus.2ex}%
      {-1em}%
      {\normalfont\normalsize\sffamily\bfseries}}
\renewcommand\subparagraph{\@startsection{subparagraph}{5}{\z@}%
      {1ex \@plus.5ex \@minus .2ex}%
      {-1em}%
      {\normalfont\normalsize\sffamily\bfseries}}
\makeatother

\def\qsqmax{q^2_\mathrm{max}}

\def\modvub{|V_{ub}|}
\def\n#1e#2n{#1\times10^{#2}}

\def\gev{\,\mathrm{GeV}}

\begin{document}
%% \begin{flushright}
%% SHEP--????
%% \end{flushright}

\begin{center}\Large\bfseries\sffamily
$\modvub$ from Exclusive Semileptonic $B\to\rho$ Decays
\end{center}

\begin{center}
\textbf{\textsf{Jonathan M Flynn${}^\mathrm{a}$, Yoshiyuki
  Nakagawa${}^\mathrm{b}$, Juan
  Nieves${}^\mathrm{c}$ and Hiroshi Toki${}^\mathrm{b}$}}\\[2ex]
${}^\mathrm{a}$School of Physics and Astronomy, University of
  Southampton, Southampton SO17~1BJ, UK\\
${}^\mathrm{b}$Research Center for Nuclear Physics, Osaka
  University,
  Ibaraki, Osaka 567-0047, Japan \\
${}^\mathrm{c}$Instituto de F{\'\i}sica Corpuscular (centro mixto
CSIC-UV),
Institutos de Investigaci\'on de Paterna,
46071, Valencia, Spain
\end{center}
\medskip

\begin{quote}
\begin{center}\textbf{\textsf{Abstract}}\end{center}
We use Omn\`es representations of the form factors $V$, $A_1$ and
$A_2$ for exclusive semileptonic $B\to\rho$ decays, and apply them to
combine experimental partial branching fraction information with
theoretical calculations of the three form factors to extract
$\modvub$. We find a slightly lower result,
$\modvub=(2.8\pm0.2)\times10^{-3}$ , than the values extracted from
exclusive semileptonic $B\to\pi$ decays,
$(3.47\pm0.29\pm0.03)\times10^{-3}$~\cite{Flynn:2007ii},
$(3.36\pm0.23)\times10^{-3}$~\cite{Bourrely:2008za},
$(3.38\pm0.35)\times10^{-3}$~\cite{Bailey:2008wp}, and using all other
inputs in CKM fits, $(3.55\pm0.15)\times10^{-3}$~\cite{Bona:2006ah,
UTfit:web}.  The disagreement is greater when we compare to the result
extracted from inclusive $B \to X_u l \nu$ decays, $\modvub =
(4.10\pm0.30_\mathrm{exp}\pm0.29_\mathrm{th})
\times10^{-3}$~\cite{Neubert:FPCP2007}.

\end{quote}

\section{Introduction}

The magnitude of the element $V_{ub}$ of the Cabibbo-Kobayashi-Maskawa
(CKM) quark mixing matrix plays a critical role in testing the
consistency of the Standard Model of particle physics and, in
particular, the description of CP violation. Any inconsistency could
be a sign of new physics beyond the standard model. $V_{ub}$ is
currently the least well-known element of the CKM matrix and
improvement in the precision of its determination is highly desirable
and topical.

$\modvub$ can be determined using inclusive or exclusive charmless
semileptonic $B$ decays. The inclusive method has historically
provided a more precise result, but recent
experimental~\cite{Athar:2003yg,Aubert:2005cd,Hokuue:2006nr,Aubert:2006ry,Gray:2007pw,
Adam:2007pv} and theoretical
developments~\cite{Arnesen:2005ez,Becher:2005bg,LCSR_04_BZ,
Flynn:2000gd, Flynn:2006vr,Ball:2006jz, Flynn:2007ki,
Flynn:2007qd,Bourrely:2008za, Duplancic:2008ix, Duplancic:2008zz} are
allowing the exclusive semileptonic $B\to \pi$ method to approach the
same level of precision.

Recently~\cite{Flynn:2007ii} we extracted $\modvub$ from combined
experimental partial branching fraction information and theoretical
[lattice QCD (LQCD) and Light cone sum rules (LCSR)] information on
exclusive semileptonic $B\to\pi$ decays. The Omn\`es representation
was employed to provide parametrisations of the form factors. The
extracted value turned out to be in striking agreement with that
extracted using all other inputs in CKM fits and in some disagreement
with $\modvub$ extracted from inclusive semileptonic decays.

The aim of this letter is to extend the above formalism to study the
exclusive semileptonic $B\to \rho$ decay and independently extract
$\modvub$ from the recent measurements of the partially integrated
branching fraction by BABAR~\cite{Aubert:2005cd},
Belle~\cite{Hokuue:2006nr} and CLEO~\cite{Gray:2007pw,Adam:2007pv}. We
will make use of quenched LQCD form factor
results~\cite{BowlerKC:JHEP05:2004, Abada:2002ie} for the high $q^2$
region, and LCSR values~\cite{Ball:2004rg} at $q^2=0$. Thanks to the
Omn\`es representation of the form-factors, we are able to combine all
these inputs, as we previously showed for $B\to\pi$ decays.

\section{Fit Procedure}

\subsection{Form-factors and differential decay width}

The semileptonic decay $B^0 \to \rho^- \ell^+ \nu_l$ is determined by
the matrix element of the $V-A$ weak current between a $B$ meson and a
$\rho$ meson. The matrix element is
\begin{equation}
\langle \rho (k, \eta) | \bar b \gamma^\mu (1-\gamma_5) u | B
(p) \rangle = \eta^*_\beta T^{\mu\beta}, 
\end{equation}
with form factor decomposition
\begin{eqnarray}
T_{\mu\beta} &=&
\frac{2V(q^2)}{m_B+m_\rho}\epsilon_{\mu\gamma\delta\beta}p^\gamma
k^\delta - \mathrm{i} (m_B+m_\rho)A_1(q^2)
g_{\mu\beta}\nonumber\\ &&\mbox{} + \mathrm{i}
\frac{A_2(q^2)}{m_B+m_\rho} (p+k)_\mu q_\beta - \mathrm{i}
\frac{2A(q^2)}{q^2} m_\rho q_\mu (p+k)_\beta,
\end{eqnarray}
where $q=p-k$ is the four-momentum transfer and $\eta$ is the $\rho$
polarisation vector. The meson masses are $m_B=5279.5$ MeV and
$m_\rho=775.5$ MeV for $B^0$ and $\rho^-$, respectively. In the
helicity basis each of the form factors corresponds to a transition
amplitude with definite spin-parity quantum numbers in  the center
of mass frame of the lepton pair. This relates the form factors $V$,
$A_1$ and $A_2$ to the total angular momentum and parity quantum
numbers of the $B\rho$ meson pair, $J^P=1^-, 1^+$ and $1^+$,
respectively~\cite{Wirbel:1985ji}. The physical region for the squared
four-momentum transfer is $0\le q^2 \le \qsqmax \equiv
(m_B-m_\rho)^2$. If the lepton mass can be ignored ($l=e$ or $\mu$),
the total decay rate is given by
\begin{equation}
\Gamma\left( B^0 \to \rho^- \ell^+ \nu_l \right) =
\frac{G_F^2|V_{ub}|^2}{192\pi^3m^3_B} \int_0^{q^2_{\rm max}}
dq^2 q^2\left[\lambda (q^2)\right]^\frac12 \left
(|H^+(q^2)|^2+|H^-(q^2)|^2+|H^0(q^2)|^2  \right)
\label{eq:gamma}
\end{equation}
where $G_F= 1.16637\times 10^{-5}$ GeV$^{-2}$ is the Fermi constant
and $\lambda(q^2)=(m^2_B+m^2_\rho-q^2)^2-4m^2_Bm^2_\rho$. $H^0$ comes
from the contribution of the longitudinally polarised $\rho$ and is
given by 
\begin{equation}
H^0(q^2) = - \frac{1}{2m_\rho\sqrt {q^2}}\left \{ \left(
m_B^2-m^2_\rho-q^2\right)\left( m_B+m_\rho\right)A_1(q^2)-
\frac{4m_B^2|\vec{k}\,|^2}{m_B+m_\rho}A_2(q^2) \right \}
\end{equation}
where $\vec{k}$ is the momentum of the $\rho$ in the $B$-meson rest
frame. $H^{\pm}$ correspond to the contribution of the transverse
 polarisations of the vector meson and are given by~\footnote{Note a
   typo in Eq.~(1.7) of Ref.~\cite{BowlerKC:JHEP05:2004}, the $\pm$
   sign should be $\mp$, as used in previous papers of the UKQCD
   Collaboration~\cite{Bowler:1994zr, Flynn:1995dc}.}
\begin{equation}
H^{\pm} = - \left \{ (m_B+m_\rho) A_1(q^2) \mp
\frac{2m_B|\vec{k}\,|}{m_B+m_\rho} V(q^2) \right \}
\end{equation}
The CLEO Collaboration has also measured partial branching fractions
of the differential distribution~\cite{Gray:2007pw, Adam:2007pv}
\begin{eqnarray}
\frac{d\Gamma(B^0\to \rho^- \ell^+\nu)}{dq^2\,d\cos\theta_{W\ell}} & =
   &  
   \frac{G_F^2 |V_{ub}|^2}{512\pi^3 m_B^3} q^2 \left [ \lambda
   (q^2)\right]^\frac12 \Big\{ 2\sin^2\theta_{W\ell} |H^0(q^2)|^2
        \nonumber \\ 
 & & \mbox{}  +
      (1-\cos\theta_{W\ell})^2|H^+(q^2)|^2+ 
(1+\cos\theta_{W\ell})^2|H^-(q^2)|^2 
              \Big\} \label{eq:cos}
\end{eqnarray}
with $\theta_{W\ell}$ the angle between the charged lepton direction in
the virtual $W-$gauge boson rest frame and the virtual $W$ in the
$B$-meson rest frame.

\subsection{Omn\`es parametrisations}

We have
previously~\cite{Flynn:2007ii,Flynn:2000gd,Flynn:2006vr,Flynn:2007ki,
  Flynn:2007qd } used a multiply subtracted Omn\`es dispersion
relation~\cite{omnes,mushkelishvili}, based on unitarity and
analyticity properties, to describe $B \to \pi$ semileptonic decays.
Here, we apply these ideas to $B \to \rho$ decays and use for $(n+1)$
subtractions~\cite{Flynn:2007qd}
\begin{equation}
F(q^2) = \frac{1}{s_0-q^2} \prod_{i=0}^n \left[F
  (s_i)(s_0-s_i)\right]^{\alpha_i(q^2)}, \quad  \alpha_i(s)\equiv
  \prod_{j=0, j\ne i} \frac{s-s_j}{s_i-s_j}, \qquad F=V,A_1,A_2
  \label{eq:omn-param}
\end{equation}
where $s_0$ corresponds to a pole of the form factor $F$. We fix
$s_0=m_{B^*}^2$ and $s_0 = s_{\rm th}=(m_B+m_\rho)^2$ for $V$ and
$A_1$ and $A_2$ form factors, respectively. In principle, for the
axial form factors one should use the square of the $1^+$ $B$-meson
mass. The mass of this latter hadron is not well established yet, but
it appears to be heavier than the $1^-$ $B^*$ resonance. Thus and for
the purposes of this exploratory work, since it would be reasonably
far from $\sqrt{q^2_{\rm max}}$, it is sufficient to employ $s_{\rm
  th}$. The parametrisation of Eq.~(\ref{eq:omn-param}) amounts to
finding an interpolating polynomial for $\ln[(s_0-q^2)F(q^2)]$ passing
through the points $(s_0^2-s_i)F(s_i)$. While one could always propose
a parametrisation using an interpolating polynomial for
$\ln[g(q^2)F(q^2)]$ for a suitable function $g(q^2)$, the derivation
using the Omn\`es representation shows that taking $g(q^2)=s_0^2-q^2$
is physically motivated~\cite{Flynn:2007qd}.

%
%%%%%%%%%%%%%%%%%%%%%%%%%%%%%%%%%%%%%%%%%%%%%%%%%%%%%%%%
% Input data
%%%%%%%%%%%%%%%%%%%%%%%%%%%%%%%%%%%%%%%%%%%%%%%%%%%%%%%%
\subsection{Theoretical and experimental inputs}

We have used experimental partial branching fraction data from
CLEO~\cite{Gray:2007pw,Adam:2007pv}, Belle~\cite{Hokuue:2006nr} and
BABAR~\cite{Aubert:2005cd}. CLEO and BABAR combine results for neutral
and charged $B$-meson decays using isospin symmetry, while Belle give
separate values for $B^0 \to \rho^- \ell^+ \nu_l$ and $B^+ \to \rho^0
\ell^+ \nu_l$ decays. Belle use three $q^2$ intervals, and we have
added in quadrature the two different systematic errors quoted for
each $q^2$ bin, and combined charged and neutral $B$-meson results. We
take the resulting systematic errors to be fully correlated. BABAR's
untagged analysis also uses three $q^2$ bins and we have assumed that
the quoted percentage systematic errors for the partial branching
fractions divided by total branching fraction are representative for
the partial branching fractions alone and, following BABAR, took them
to be fully correlated. CLEO determines partial branching fractions as
a function of both $q^2$ and of $\cos\theta_{W\ell}$ (see
Eq.~(\ref{eq:cos})) and complete correlation matrices are given
in~\cite{Gray:2007pw} for both statistical uncertainties and
systematic errors that we have used in our fits.

When computing partial branching fractions, we have used $\tau_{B^0}=
1/\Gamma_\mathrm{Tot} = \n(1.527\pm
0.008)e-12n\,\mathrm{s}$~\cite{Barberio:2007cr} for the $B^0$
lifetime. All the branching fraction inputs are listed in
Table~\ref{tab:in_exp}.

\begin{table}
\begin{center}
\begin{tabular}{ @{} l >{$} c <{$} >{$} c <{$} >{$} c <{$} >{$} c <{$} @{}  }
\hline
 & q^2 \mbox{\ range } [\gev^2] & \cos\theta_{W\ell} \mbox{\ range }
 & 10^4{B}^{\textrm{in}}_k & 10^4{B}^{\textrm{Omn\`es}}_k \\
 \hline
BELLE~\cite{Hokuue:2006nr}
  & 0 -  8 & [-1,1] & 0.62 \pm 0.14 \pm 0.06 & 0.69 \pm 0.12\\
  & 8 - 16 & [-1,1] & 1.20 \pm 0.23 \pm 0.11 & 1.12 \pm 0.15\\
  & > 16   & [-1,1] & 0.53 \pm 0.20 \pm 0.12 & 0.53 \pm 0.08\\
\hline
BABAR~\cite{Aubert:2005cd}
  & 0 - 10  & [-1,1] & 0.73 \pm 0.17 \pm 0.21 & 0.96 \pm 0.15\\
  & 10 - 15 & [-1,1] & 0.82 \pm 0.10 \pm 0.13 & 0.71 \pm 0.10\\
  & > 15    & [-1,1] & 0.59 \pm 0.07 \pm 0.16 & 0.68 \pm 0.10\\
\hline
CLEO ~\cite{Gray:2007pw}
 &  0 -  2 & [-1,1] & 0.45 \pm 0.20 \pm 0.15 & 0.08 \pm 0.03\\
 &  2 -  8 & [-1,1] & 0.96 \pm 0.20 \pm 0.29 & 0.61 \pm 0.10\\
 &  8 - 16 & \phantom{-}[0,1] & 0.75 \pm 0.16 \pm 0.14 & 0.74 \pm 0.10\\
 & > 16    & \phantom{-}[0,1] & 0.35 \pm 0.07 \pm 0.05 & 0.39 \pm 0.06\\
 & >  8    & [-1,0] & 0.42 \pm 0.18 \pm 0.31 & 0.51 \pm 0.07\\
\hline
\end{tabular}
\end{center}
\caption{Experimental branching fraction inputs for the $\chi^2$
  function defined in Eq.~(\ref{eq:chi2}). Statistical and systematic
  errors are shown. We also give branching fractions calculated using
  our fitted form factors and $|V_{ub}|$.}
\label{tab:in_exp}
\end{table}
 For theoretical form-factor inputs (listed in Table~\ref{tab:in_th}),
 we use the lightcone sumrule (LCSR) results at $q^2=0$ of
 Ref.~\cite{Ball:2004rg} and lattice QCD results from the
 UKQCD~\cite{BowlerKC:JHEP05:2004} and SPQcdR~\cite{Abada:2002ie}
 Collaborations, near $\qsqmax$. LQCD inputs have been obtained in the
 quenched approximation. There is therefore an uncontrolled systematic
 error, which is not fully included in the errors given in
 Table~\ref{tab:in_th}. 
\begin{table}
\begin{center}
\begin{tabular}{@{}l >{$}c<{$} >{$}c<{$} >{$}c<{$} >{$}c<{$} @{}}
\hline
 & q^2 \mbox{ [GeV$^2$]} & V   & A_1             & A_2 \\
 \hline
LCSR~\cite{Ball:2004rg} &     0 & 0.323 \pm 0.029 & 0.242 \pm 0.024 & 0.221 \pm 0.023 \\
 \hline \tstrut
UKQCD~\cite{BowlerKC:JHEP05:2004} 
     & 12.67 & 0.684 \pm 0.162 \, ^{+ 0.00} _{- 0.56 }& 0.439 \pm 0.067 \,^{+ 0.000} _{-0.080 } & 0.70 \pm 0.49\,^{+ 0.08 } _{- 0.03 }\\\tstrut
     & 13.01 & 0.714 \pm 0.162 \, ^{+ 0.00} _{- 0.50} & 0.448 \pm 0.065  \,^{+ 0.000} _{-0.079 }& 0.71 \pm 0.46\,  ^{+ 0.08 } _{- 0.03 }\\\tstrut
     & 13.51 & 0.763 \pm 0.155 \,^{+ 0.00} _{- 0.40 } & 0.460 \pm 0.063  \,^{+ 0.000} _{-0.075 }& 0.72 \pm 0.43\, ^{+ 0.10 } _{- 0.02 }\\\tstrut
     & 14.02 & 0.818 \pm 0.147 \, ^{+ 0.00} _{- 0.31 }& 0.472 \pm 0.059  \,^{+ 0.000} _{-0.073 }& 0.73 \pm 0.42\, ^{+ 0.12 } _{- 0.01 } \\\tstrut
     & 14.52 & 0.883 \pm 0.141 \, ^{+ 0.00} _{- 0.24 }& 0.485 \pm 0.055  \, ^{+ 0.000} _{-0.070 }& 0.76 \pm 0.42\,^{+ 0.14 } _{- 0.03 } \\\tstrut
     & 15.03 & 0.967 \pm 0.137 \, ^{+ 0.00} _{- 0.20} & 0.498 \pm 0.051  \,^{+ 0.000} _{-0.068 }& 0.78 \pm 0.46\,^{+ 0.16 } _{- 0.05 } \\\tstrut
     & 15.53 & 1.057 \pm 0.134 \, ^{+ 0.00} _{- 0.19} & 0.513 \pm 0.049  \, ^{+ 0.000} _{-0.067 }& 0.81 \pm 0.54\,^{+ 0.18} _{- 0.06 } \\\tstrut
     & 16.04 & 1.164 \pm 0.150 \,^{+ 0.10 } _{- 0.21}  & 0.529 \pm 0.047  \,^{+ 0.000} _{-0.066 }& 0.84 \pm 0.71\,^{+ 0.20 } _{- 0.07 } \\\tstrut
     & 16.54 & 1.296 \pm 0.184 \,^{+ 0.21 } _{- 0.25}  & 0.544 \pm 0.043  \,^{+ 0.000} _{-0.062 }& 0.87 \pm 0.97\,^{+ 0.23 } _{- 0.08 } \\\tstrut
     & 17.05 & 1.46 \pm 0.26   \,^{+ 0.34 } _{- 0.30} & 0.560 \pm 0.043  \,^{+ 0.000} _{-0.059 }& 0.90 \pm 1.35\,^{+ 0.27 } _{- 0.07 } \\\tstrut
     & 17.55 & 1.67 \pm 0.40   \, ^{+ 0.49 } _{- 0.36}& 0.577 \pm 0.043  \,^{+ 0.000} _{-0.058 }& 0.90 \pm 1.89\,^{+ 0.33 } _{- 0.03 } \\\tstrut
     & 18.17 & 2.02 \pm 0.68   \, ^{+ 0.73 } _{- 0.48} & 0.599 \pm
 0.052  \,^{+ 0.000 } _{-0.058 } & 0.9 \pm 2.9\,^{+ 0.4 } _{- 0.1 } \\\\
\hline\tstrut
SPQcdR~\cite{Abada:2002ie}
      & 10.69 & 0.51 \pm 0.26  & 0.354 \pm 0.085 &  0.38 \pm 0.26\\
      &  12.02 & 0.61 \pm 0.28 & 0.384 \pm 0.087 &  0.49 \pm 0.30\\
      &  13.35 & 0.74 \pm 0.30 & 0.421 \pm 0.089 &  0.65 \pm 0.35\\
      &  14.68 & 0.93 \pm 0.31 & 0.465 \pm 0.092  &  0.93 \pm 0.41\\
      &  16.01 & 1.20  \pm 0.32& 0.519 \pm 0.097 &  1.41  \pm 0.56\\
      &  17.34 & 1.61  \pm 0.33& 0.588 \pm 0.108  &  2.39 \pm 1.23\\
      &  18.67 & 2.26  \pm 0.55& 0.678 \pm 0.134  &  4.7 \pm 4.1\\
\hline
\end{tabular}
\end{center}
\caption{Form factor inputs for the $\chi^2$ function defined in
  Eq.~(\ref{eq:chi2}). For UKQCD we show both statistical (symmetrized)
  and systematical errors, while  SPQcdR errors include both
  systematic and statistical uncertainties (we are indebted with
  C.M. Maynard and F. Mescia for providing us with these form factors).}
\label{tab:in_th}
\end{table}

\subsection{Definition of $\chi^2$}

We implement the following fitting procedure. Choose a set of
subtraction points spanning the physical range to use in the Omn\`es
formula of equation~(\ref{eq:omn-param}). Now find the best-fit value
of $\modvub$ and the form factors at the subtraction points to match 
both theoretical input form factor values and the experimental partial
branching fraction inputs. The $\chi^2$ function for the fit is :
\begin{eqnarray}
\chi^2 &=& \sum_{i,j=1}^{60}
 \left[F^\mathrm{in}_i-\fom{i}\right]
  C^{-1}_{ij}\left[F^\mathrm{in}_j-\fom{j}\right]\nonumber\\
  & & \mbox{} + 
 \sum_{k,l=1}^{11}
  \left[B_k^\mathrm{in} -
   B_k^{\mbox{\scriptsize Omn\`es}}(\modvub,F_0,F_1,F_2)\right]
   C^{-1}_{B\,kl}
   \left[B_l^\mathrm{in} -
   B_l^{\mbox{\scriptsize Omn\`es}}(\modvub,F_0,F_1,F_2)\right],
   \label{eq:chi2}
\end{eqnarray}
where $F^\mathrm{in}_i$ are input LCSR or lattice QCD values for
$V(q^2_i), \,A_1(q^2_i)$ and $A_2(q^2_i)$, and $B^\mathrm{in}_k$ are
input experimental partial branching fractions. Moreover, $\fom i$
stands for each of the form factors $F=V,A_1, A_2$ at $q^2=q^2_i$, and
it is given by equation~(\ref{eq:omn-param}) with three subtractions
$(s_l, F(s_l))$ at $(0,F_0)$, $(2\qsqmax/3,F_1)$ and
$(\qsqmax,F_2)$. The branching fractions $B^{\mbox{\scriptsize
Omn\`es}}$ are calculated using $F^{\mbox{\scriptsize Omn\`es}}$, for
$V, A_1$ and $A_2$ form factors . There are in total 10 fit
parameters: $V(0)$, $V(2\qsqmax/3)$, $V(\qsqmax)$, $A_1(0)$,
$A_1(2\qsqmax/3)$, $A_1(\qsqmax)$, $A_2(0)$, $A_2(2\qsqmax/3)$,
$A_2(\qsqmax)$ and $\modvub$. The latter parameter is used when
computing $B^{\mbox{\scriptsize Omn\`es}}$.

We have assumed that the LCSR and LQCD form factor values have
independent statistical uncertainties and treated the errors listed in
Table~\ref{tab:in_th} for the SPQcdR inputs as purely statistical. For
the UKQCD data we have put the form factor values in the centre of
their systematic range and use half that range as the systematic
error. We have built a covariance matrix where the statistical
uncertainties ($\sigma_i$) are uncorrelated and the systematic errors
($\epsilon_i$) are fully correlated, leading to a $60\times 60$
covariance matrix with three diagonal blocks. The first $3\times 3$
and second $21\times 21$ blocks are for the LCSR and SPQcdR results
and have the form $C_{ij}=\sigma_i^2 \delta_{ij}$. The third block is
for the UKQCD data and has the form $C_{ij}=\sigma_i^2
\delta_{ij}+\epsilon_i\epsilon_j$. We will further discuss the effect
of the UKQCD systematic errors on $\modvub$ below.
 
The covariance matrix, $C_B$, for the partial branching
fraction inputs is constructed as follows. For Belle and BABAR input
data, we have assumed independent statistical uncertainties and
fully-correlated systematic errors leading to an
$6\times 6$ covariance matrix with two diagonal blocks of the form
$C_{B\,ij}=\sigma_i^2 \delta_{ij} + \epsilon_i\epsilon_j$. For the
CLEO input, we use an $5\times 5$ covariance matrix $C_{B\,ij}^{\rm
CLEO}= \sigma_i \sigma_j {\cal C}_{B\,ij}^{\,\rm CLEO-stat}+
\epsilon_i \epsilon_j {\cal C}_{B\,ij}^{\,\rm CLEO-sys} $, where we
have read off the statistical and systematic correlation matrices
(${\cal C}_{B\,ij}^{\,\rm CLEO-stat/sys}$) from
tables X and XI, respectively, of Ref.~\cite{Gray:2007pw}. 

We do not consider any correlation between measurements from different
experiments, or between different sources of theoretical inputs. Nor
do we consider correlations between experimental and theoretical
inputs.

%%%%%%%%%%%%%%%%%%%%%%%%%%%%%%%%%%%%%%%%%%%%%%%%%%%%%%%%
% Results
%%%%%%%%%%%%%%%%%%%%%%%%%%%%%%%%%%%%%%%%%%%%%%%%%%%%%%%%
\section{Results and discussion}

The best fit parameters are 
\begin{equation}
\begin{array}{rcl}
  \modvub         & = & \left (2.76  \pm 0.21\right) \times 10^{-3} \\
  V(0)            & = & 0.322 \pm 0.030 \\
  V(2\qsqmax/3)   & = & 0.681 \pm 0.073 \\
  V(\qsqmax)      & = & 4.21  \pm 0.76  \\
  A_1(0)          & = & 0.223 \pm 0.021 \\
  A_1(2\qsqmax/3) & = & 0.449 \pm 0.020 \\
  A_1(\qsqmax)    & = & 0.657 \pm 0.055 \\
  A_2(0)          & = & 0.231 \pm 0.022 \\
  A_2(2\qsqmax/3) & = & 0.679 \pm 0.098 \\
  A_2(\qsqmax)    & = & 2.76  \pm 1.38  \\
\end{array}
\label{eq:besfit}
\end{equation}
The fit has $\chi^2/\mathrm{d.o.f.}=0.21$ for $61$ degrees of freedom,
while the Gaussian correlation matrix can be found in the
appendix~\ref{sec:app}.  In figure~\ref{fig:results} we show the
fitted form factors and the differential decay rate calculated from
our fit.  Partial branching fractions calculated for the same bins as
used experimentally are given in the last column of
Table~\ref{tab:in_exp}. Our calculated total branching ratio turns out
to be $(2.30^{+0.24}_{-0.26}) \times 10^{-4}$, in reasonable agreement
with $(2.80\pm 0.18 \pm 0.16)\times 10^{-4}$ quoted by the Heavy
Flavours Averaging Group (HFAG)~\cite{Barberio:2007cr}.

We have further investigated the effect of the highly asymmetric UKQCD
systematic errors on $\modvub$. First, we have completely dropped them
and used only the statistical uncertainties on the UKQCD points. We
find $\modvub= \left (2.68 \pm 0.19\right) \times 10^{-3}$. Second, we
have performed a Monte Carlo where we randomly choose each UKQCD form
factor value within its systematic error range, with complete
correlation between all systematic shifts. For each trial we perform a
fit like our original one, but setting to zero the systematic errors
on the UKQCD inputs. In this case, we find $\modvub= \left (2.85 \pm
0.10\right) \times 10^{-3}$. Note that this last result is the mean
and the standard deviation of the fit result for $\modvub$ over all
the trials, whereas the result above and that quoted in
Eq.~(\ref{eq:besfit}) are the fit result and error from a single fit.
Thus, the result from this second procedure should be understood as a
shift of $+0.09 \pm 0.10$ in the value of $\modvub$ in
Eq.~(\ref{eq:besfit}). Finally, we have repeated the latter procedure,
but taking the $A_2$ systematic error to be anticorrelated with those
of the $V$ and $A_1$. This results in $\modvub= \left (2.86 \pm
0.15\right) \times 10^{-3}$.

From the above discussion, we estimate
\begin{equation}
\modvub = 2.8 \pm 0.2
\label{eq:vubresult}
\end{equation}
which constitutes our main result. Quenched approximation systematic
effects from LQCD are not accounted for by the $0.2$ error quoted
above. These are difficult to quantify and are a limitation here.
However, unquenched lattice simulations are now standard and future
lattice QCD results will address this limitation (although they will
also face the problem of an unstable $\rho$ meson for light enough
simulated up and down quark masses).
\begin{figure}
\begin{center}
\includegraphics[width=\textwidth]{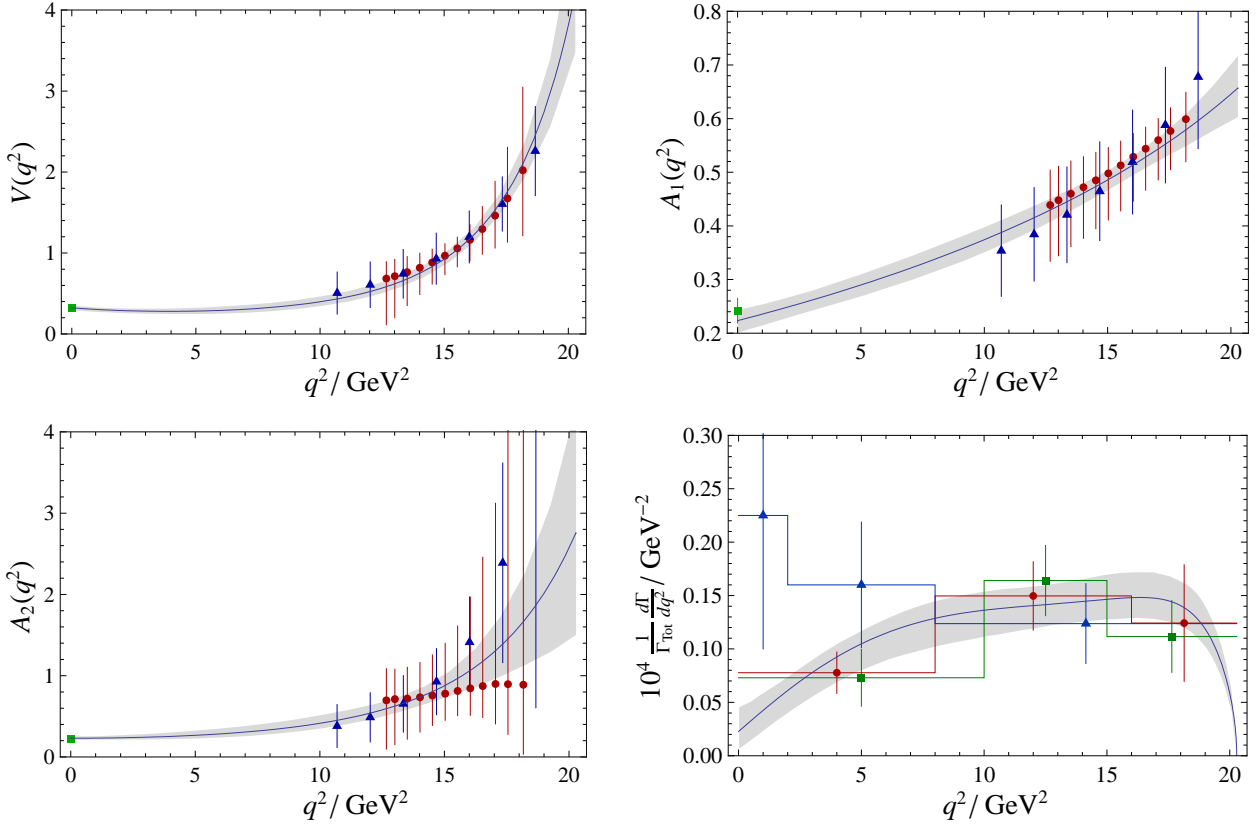}
\end{center}
\caption{Results obtained from the fit to experimental partial
  branching fraction data and theoretical form factor calculations.
  The top and the left bottom plots show the three form factors with
  their 68\% CL  bands (shaded) together with
  the lattice and LCSR input points (green square LCSR, red dots
  UKQCD, blue triangles
  SPQcdR).  The bottom right plot shows the differential decay rate  with
  68\% CL  band (shaded)  together
  with the experimental partial branching fractions divided by the
  appropriate bin-width (histograms and points). Green squares, red
  dots and blue triangles
  denote BABAR, Belle and CLEO results, respectively.}
\label{fig:results}
\end{figure}
Nevertherless, we see that the Omn\`es framework used here provides a
fair description of all available experimental and theoretical results
for semileptonic $B\to\rho$ decays, leading to a further independent
determination of $\modvub$. The result is lower than the values
obtained in the most recent studies of the exclusive semileptonic
$B\to\pi$ decay,
$(3.47\pm0.29\pm0.03)\times10^{-3}$~\cite{Flynn:2007ii},
$(3.36\pm0.23)\times10^{-3}$~\cite{Bourrely:2008za},
$(3.38\pm0.35)\times10^{-3}$~\cite{Bailey:2008wp}, $(3.5 \pm
0.4_\mathrm{th} \pm 0.2_\mathrm{shape}\pm
0.1_\mathrm{BR})\times10^{-3}$~\cite{Duplancic:2008ix}, and using all
other inputs in CKM fits,
$(3.55\pm0.15)\times10^{-3}$~\cite{Bona:2006ah, UTfit:web}. The
disagreement is greater when we compare to the most precise result
extracted from inclusive $B \to X_u l \nu$ decays, $\modvub =
(4.10\pm0.30_\mathrm{exp}\pm0.29_\mathrm{th})
\times10^{-3}$~\cite{Neubert:FPCP2007}. Thus, the hints of
disagreement between inclusive and exclusive/global-CKM-fit
determinations are strengthened.

\subsubsection*{Acknowledgments}

JMF and JN acknowledge support from the EU Human Resources and
Mobility Activity, FLAVIAnet, contract number MRTN--CT--2006--035482,
PPARC grant PP/D000211/1 and MEC grant FIS2005--00810.

\appendix
\section{Gaussian correlation matrix}
\label{sec:app}
Here, we give the correlation matrix of fitted parameters
corresponding to the best-fit parameters in Eq.~(\ref{eq:besfit})
\begin{equation}
\left(
\begin{array}{rrrrrrrrrr}
   1.00 & -0.01 & -0.21 & -0.16 & -0.10 & -0.31 & -0.23 &  0.05 &  0.34 &  0.01 \\
        &  1.00 &  0.01 &  0.04 &  0.00 &  0.00 &  0.00 &  0.00 &  0.00 & -0.01 \\
        &       &  1.00 & -0.44 &  0.05 &  0.28 &  0.21 & -0.03 &  0.16 &  0.07 \\
        &       &       &  1.00 & -0.05 &  0.09 & -0.08 &  0.02 & -0.08 &  0.05 \\
        &       &       &       &  1.00 &  0.10 & 0.03  &  0.17 &  0.12 & -0.46 \\
        &       &       &       &       &  1.00 & -0.32 & -0.04 &  0.34 & -0.30 \\
        &       &       &       &       &       &  1.00 &  0.04 & -0.27 &  0.47 \\
        &       &       &       &       &       &       &  1.00 & -0.04 &  0.21 \\
        &       &       &       &       &       &       &       &  1.00 &  0.18 \\
        &       &       &       &       &       &       &       &       &  1.00 \\
\end{array}
\right)
\end{equation}

%%%%%%%%%%%%%%%%%%%%%%%%%%%%%%%%%%%%%%%%%%%%%%%%%%%%%%%%
% Bibliography
%%%%%%%%%%%%%%%%%%%%%%%%%%%%%%%%%%%%%%%%%%%%%%%%%%%%%%%%
\bibliographystyle{h-physrev4}

\end{document}